\documentclass[12pt]{iopart}

\usepackage{graphicx} 

\begin{document}

\title[]{Universal growth of social groups: empirical analysis and modeling}

\author{Ana Vrani\'c}
\address{Institute  of  Physics  Belgrade, University of Belgrade, Pregrevica 118, 11080 Belgrade, Serbia}
\ead{ana.vranic@ipb.ac.rs}

\author{Jelena Smiljani\'c}
\address{Integrated Science Lab, Department of Physics, Umeå University, SE-901 87 Umeå, Sweden\\
Institute  of  Physics  Belgrade, University of Belgrade, Pregrevica 118, 11080 Belgrade, Serbia}
\ead{jelena.smiljanic@ipb.ac.rs}

\author{Marija Mitrovi\'c Dankulov}  
\address{Institute  of  Physics  Belgrade, University of Belgrade, Pregrevica 118, 11080 Belgrade, Serbia}
\ead{marija.mitrovic.dankulov@ipb.ac.rs}

\begin{abstract}
Social groups are fundamental elements of any social system. Their emergence and evolution are closely related to the structure and dynamics of a social system. Research on social groups was primarily focused on the growth and the structure of the interaction networks of social system members and how members' group affiliation influences the evolution of these networks. The distribution of groups' size and how members join groups has not been investigated in detail. Here we combine statistical physics and complex network theory tools to analyze the distribution of group sizes in three data sets, Meetup groups based in London and New York and Reddit. We show that all three distributions exhibit log-normal behavior that indicates universal growth patterns in these systems. We propose a theoretical model that combines social and random diffusion of members between groups to simulate the roles of social interactions and members' interest in the growth of social groups. The simulation results show that our model reproduces growth patterns observed in empirical data. Moreover, our analysis shows that social interactions are more critical for the diffusion of members in online groups, such as Reddit, than in offline groups, such as Meetup. This work shows that social groups follow universal growth mechanisms that need to be considered in modeling the evolution of social systems. 

\end{abstract}

%
%
%
%
%

\section{Introduction}

The need to develop methods and tools for their analysis and modeling comes with massive data sets. Methods and paradigms from statistical physics have proven to be very useful in studying the structure and dynamics of social systems \cite{castellano2009statistical}. The main argument for using statistical physics to study social systems is that they consist of many interacting elements. Due to this, they exhibit different patterns in their structure and dynamics, commonly known as \textit{collective behavior}. While various properties can characterize a social system's building units, only a few enforce collective behavior in the systems. The phenomenon is known as \textit{universality} in physics and is commonly observed in social systems such as in voting behavior \cite{chatterjee2013universality}, or scientific citations \cite{radicchi2008universality}. It indicates the existence of the universal mechanisms that govern the dynamics of the system \cite{castellano2009statistical}.\\

Social groups, informal or formal, are mesoscopic building elements of every socio-economic system that direct its emergence, evolution, and disappearance \cite{firth2013elements}. The examples span from countries, economies, and science to society. Settlements, villages, towns, and cities are formal and highly structured social groups of countries. Their organization and growth determine the functioning and sustainability of every society \cite{barthelemy2016structure}. Companies are the building blocks of an economic system, and their dynamics are essential indicators of the level of its development \cite{hidalgo2009building}. Scientific conferences, as scientific groups, enable fast dissemination of the latest results, exchange, and evaluation of ideas as well as a knowledge extension, and thus are an integral part of science \cite{smiljanic2016theoretical}. The membership of individuals in various social groups, online and offline, can be essential when it comes to the quality of their life \cite{montazeri2001anxiety, davison2000talks, cho2012tea}. Therefore, it is not surprising that the social group emergence and evolution are at the center of the attention of many researchers \cite{aral2012identifying,gonzalez2013broadcasters, torok2013opinions, yasseri2012dynamics}.\\

The availability of large-scale and long-term data on various online social groups has enabled the detailed empirical study of their dynamics. The focus was mainly on the individual groups and how structural features of social interaction influence whether individuals will join the group \cite{backstrom2006group} and remain its active members \cite{smiljanic2016theoretical, smiljanic2017associative}. The study on LiveJournal \cite{backstrom2006group} groups has shown that decision of an individual to join a social group is greatly influenced by the number of her friends in the group and the structure of their interactions. The conference attendance of scientists is mainly influenced by their connections with other scientists and their sense of belonging \cite{smiljanic2016theoretical}. The sense of belonging of an individual in social groups is achieved through two main mechanisms \cite{smiljanic2017associative}: expanding the social circle at the beginning of joining the group and strengthening the existing connections in the later phase. Analysis of the evolution of large-scale social networks has shown that edge locality plays a critical role in the growth of social networks \cite{leskovec2008microscopic}. The dynamics of social groups depend on their size \cite{palla2007quantifying}. Small groups are more cohesive with continued long-term, while large groups change their active members constantly \cite{palla2007quantifying}.
These findings help us understand the growth of a single group, the evolution of its social network, and the influence of the network structure on group growth. However, how the growth mechanisms influence the distribution of members of one social system among groups is yet to be understood.\\
 
Furthermore, it is not clear whether the growth mechanisms of social groups are universal or system-specific. The size distribution of social groups has not been extensively studied. Rare empirical evidence of the size distribution of social groups indicates that it follows power-law behavior \cite{zheleva2009co}. However, the distribution of company sizes follows log-normal behavior and remains stable over decades \cite{amaral1997scaling, stanley1996scaling}. Analysis of the cities' sizes shows that all cities' distribution also follows a log-normal distribution \cite{}. In contrast, the distribution of the largest cities resembles Zipf's distribution \cite{fazio2015pareto}.

A related question that should be addressed is whether we can create a unique yet relatively simple microscopic model that reproduces the distribution of members between groups and explains the differences observed between social systems. French economist Gibrat proposed a simple growth model to produce companies' and cities' observed log-normal size distribution. However, the analysis of the growth rate of the companies \cite{amaral1997scaling} has shown that growth mechanisms are different from those assumed by Gibrat. In addition, the analysis of the growth of the online social networks showed that the population size and spatial factors do not determine population growth, and it deviates from Gibrat's law \cite{zhu2014online}. Other mechanisms, for instance, growth through diffusion, have been used to model and predict rapid group growth \cite{kairam2012life}. However, the growth mechanisms of various social groups and the source of the scaling observed in socio-economic systems remain hidden.\\

Here we analyze the size distribution of formal social groups in three data sets: Meetup groups based in London and New York and subreddits on Reddit. We are interested in the scaling behavior of size distributions and the distribution of growth rates. Empirical analysis of the dependence of growth rates, shown in this work, indicates that growth cannot be explained through Gibrat's model. Here we contribute with a simple microscopic model that incorporates some of the findings of previous research \cite{backstrom2006group, zheleva2009co}. We show that the model can reproduce size and growth rate distributions for both studied systems. Moreover, the model is flexible and can produce a broad set of log-normal size distributions depending on the value of model parameters.\\

The paper is organized as follows: in Section \ref{sec:data} we describe the data, while in Section \ref{sec:emp} we present our empirical results. In Section \ref{sec:model} we introduce model parameter and principles. In section \ref{sec:results} we demonstrate that model can reproduce the growth of social groups in both systems and show the results for different values of model parameters. Finally, in Section \ref{sec:con}, we present concluding remarks and discuss our results. 

\section{Data \label{sec:data}}
We analyze the growth of social groups from two widely used online platforms: Reddit and Meetup. Reddit \footnote{https://www.reddit.com/} enables sharing of diverse web content, and members of this platform interact exclusively online through posts and comments. The Meetup \footnote{www.meetup.com} allows people to use online tools to organize offline meetings. The building elements of the Meetup system are topic-focused groups, such as food lovers or data science professionals. Due to their specific activity patterns - events where members meet face-to-face - Meetup groups are geographically localized, and interactions between members are primarily offline.  

We compiled the Reddit data from https://pushshift.io/. This site collects data daily and, for each month, publishes merged comments and submissions in the form of JSON files. 
Specifically, we focus on subreddits - social groups of Reddit members interested in a specific topic. 
We selected subreddits created between $2006$ and $2011$ that were active in $2017$ and followed their growth from their beginning until $2011$. The considered dataset contains $17073$ subreddits with $2 195 677$ active members, with the oldest originating from $2006$ and the youngest being from $2011$. For each post under a subreddit, we extracted the information about the member-id of the post owner, subreddit-id, and timestamp. As we are interested in the subreddits growth in the number of members, for each subreddit and member-id, we selected the timestamp when a member made a post for the first time. Finally, in the dataset, we include only subreddits active for at least two months.  

The Meetup data were downloaded in $2018$ using public API. The Meetup platform was launched in 2003, and when we accessed the data, there were more than 240 000 active groups. For each group, we extracted information about the date it had been founded, its location, and the total number of members. We focused on the groups founded in a period between $2003$ and $2017$ in big cities, London and New York, where the Meetup platform achieved considerable popularity. We considered groups active for at least two months. There were 4673 groups with 831685 members in London and 4752 groups with 1059632 members in New York. In addition, we extracted the ids of group members, the information about organized events, and which members attended these events. Based on this, we obtained the date when a member joined a group, the first time she participated in a group event. 

For all systems, we extracted the timestamp when the member joined the group. Based on this information, we can calculate the number of new members per month $N_{i}(t)$, the group size $S_{i}(t)$ at each time step, and the growth rate for each group. The time step for all three data sets is one month. The size of the group $i$ at time step $t$ is the number of members that joined that group ending with the month, i.e., $S_{i}(t)=\sum^{k=t}_{k=t_{i0}}N_{i}(t)$, where $t_{i0}$ is the time step in which the group $i$ was created. Once the member joins the group, it has an active status by default, which remains permanent. For these reasons, the size of considered groups is a non-decreasing function. The growth rate $R_i(t)$ at step $i$ is obtained as logarithm of successive sizes $R_{i}(t)= log(S_{i}(t)/S_{i}(t-1))$. 

While the forms of communication between members and activities that members engage in differ for considered systems, some common properties exist between them. Members can form new groups and join the existing ones. Furthermore, each member can belong to an unlimited number of groups. For these reasons, we can use the same methods to study and compare the formation of groups on Reddit and Meetup.

\section{Empirical analysis of social group growth \label{sec:emp}}
\begin{figure}[!ht]
    \centering
    \includegraphics[width=0.8\linewidth]{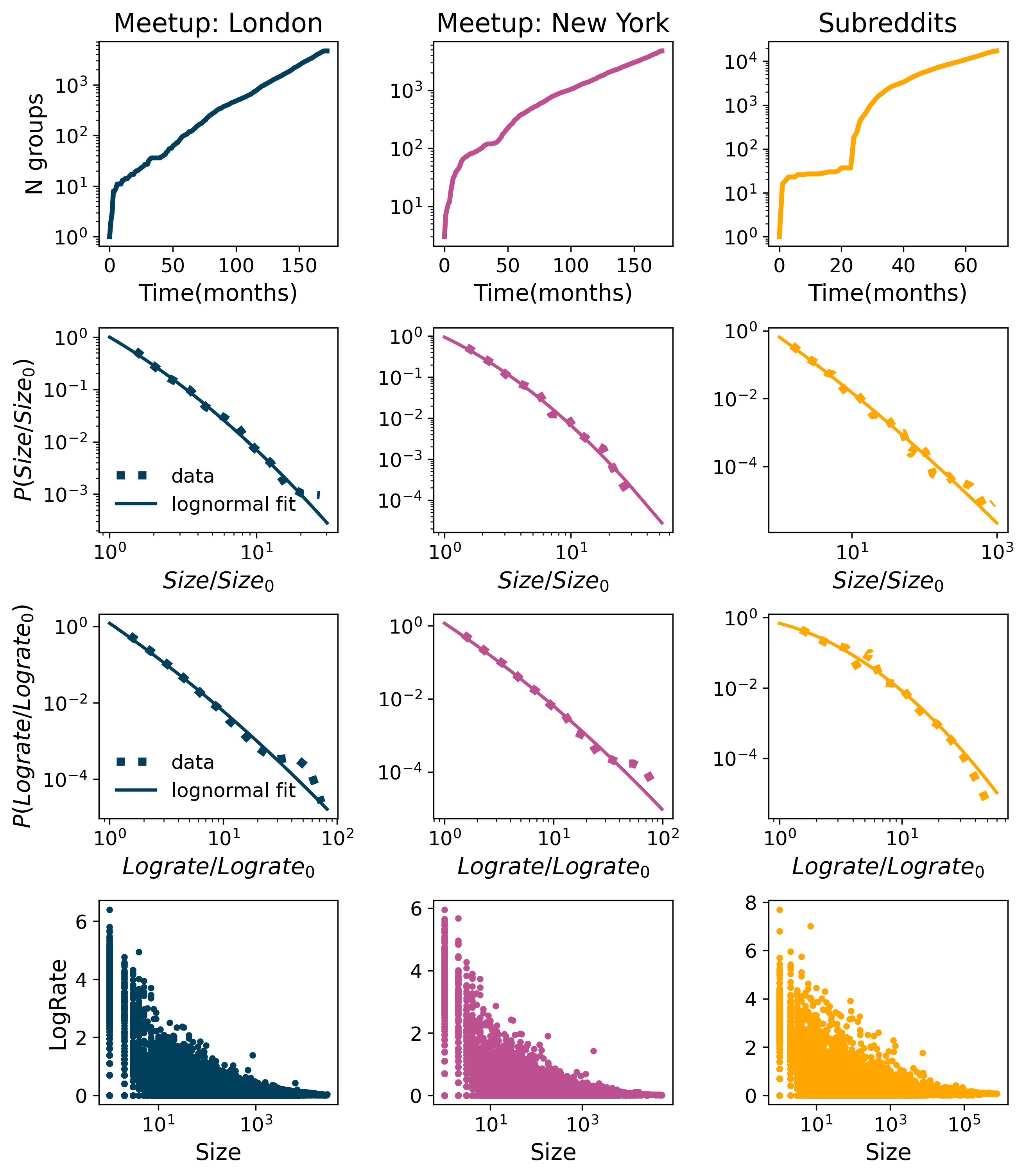}
    \caption{The number of groups over time, normalized sizes distribution, normalized log-rates distribution and dependence of log-rates and group sizes for Meetup groups created in London and New York and subreddits. The number of groups grows exponentially over time, while the group size distributions, and log-rates distributions follow log-normal. Logrates depend on the size of the group, implying that the growth cannot be explained by Gibrt law. 
    }
    \label{fig:data1}
\end{figure}

Figure \ref{fig:data1} summarizes the properties of the groups in Meetup and Reddit systems. The number of groups grows exponentially over time. Nevertheless, we notice that Reddit has a substantially larger number of groups than Meetup. The Reddit groups are prone to engage more members in a shorter period. The size of the Meetup groups ranges from several members up to several tens of thousands of members, while sizes of subreddits are between a few tens of members up to several million.
The distributions of normalized group sizes follow the log-normal distribution (see Table S1 and Fig. S1 in SI)
\begin{equation}
    P(S)=\frac{1}{{\frac{S}{S_{0}}}\sigma\sqrt{2\pi}}exp\left(-\frac{(\ln(\frac{S}{S_{0}})-\mu)^{2}}{2\sigma^{2}}\right)
    \label{eq:log} \ ,
\end{equation}
where $S$ is the group size, $S_{0}$ is the average group size in the system, and $\mu$ and $\sigma$ are parameters of the distribution. We used \textit{power-law} package \cite{powerlaw} to fit Eq. \ref{eq:log} to empirical data and found that distribution of groups sizes for Meetup groups in London and New York follow similar distributions with the values of parameters $\mu= -0.93$, $\sigma = 1.38$ and $\mu=-0.99$ and $\sigma=1.49$ for London and New York respectively. The distribution of sizes of subreddits also has the log-normal shape with parameters $\mu= -5.41$ and $\sigma = 3.07$. 

Multiplicative processes can generate the log-normal distributions \cite{mitzenmacher2004brief}. If there is a quantity with size $S_i(t)$ at time step $t$, it will grow so after time period $\delta$ the size of the quantity is $S(t+\Delta t) = S(t) r$, where $r$ represents a random number. The Gibrat law states that growth rates $r$ are uncorrelated and do not depend on the current size. To describe the growth of social groups, we calculate the logarithmic growth rates $R_{i}(t)$. According to Gibrat law the distribution of logarithmic growth rates is normal, or, as it is shown in many studies, it is better explained with Laplacian ("tent-shaped ") distribution \cite{mondani2014fat}, \cite{fu2005growth}. In Fig. \ref{fig:data1} we show the distributions of log-rates for all three data sets. Log-rates are very well approximated with a log-normal distribution. Furthermore, the bottom panel of Fig. \ref{fig:data1} shows that log-rates are not independent of group size. Figure \ref{fig:data1} shows that these findings imply that the growth of Meetup and Reddit groups violates the basic assumptions of the Gibrat's law \cite{frasco2014spatially, qian2014origin} and that it can not be explained as a simple multiplicative process.\\

\begin{figure}[!ht]
    \centering
    \includegraphics[width=0.8\linewidth]{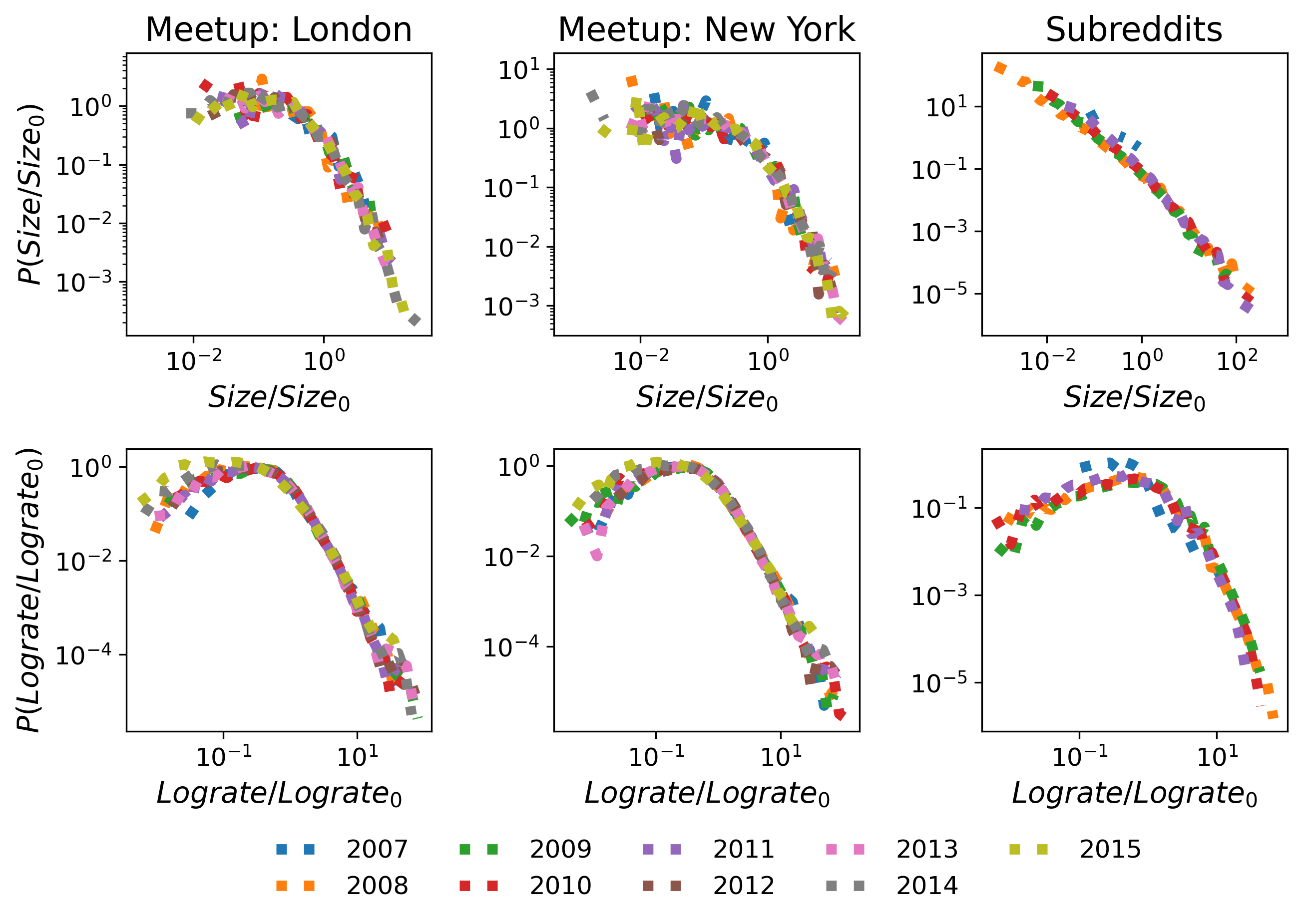}
    \caption{The figure shows the groups' sizes distributions and log-rates distributions.
    Figures in the top panel show the distribution of normalized sizes of groups created in the same year. Distributions for the same system and different years follow same log-normal distribution indicating existence of universal growth patterns. }
    \label{fig:scale}
\end{figure}

We are considering a relatively significant period for online groups. The fast expansion of information communications technologies (ICT) changed how members access online systems. With the use of smartphones, online systems became more available, which led to the exponential growth of ICTs systems and potential change in the mechanisms that influence the social groups' growth. For these reasons, we aggregate groups according to the year they were founded for each of the three data sets and look at the distributions of their sizes at the end of 2017 for Meetup groups and 2011 for Reddit. For each year and each of the three data sets, we calculate the average size of the groups created in a year $y$ $\langle S^{y}\rangle$. We normalize the size of the groups originating in year $y$ with the corresponding average size $s^{y}_{i}=S^{y}_{i}/\langle S^{y}\rangle$ and calculate the distribution of the normalized sizes for each year. The distribution of normalized sizes for all years and data sets is shown in Fig. \ref{fig:scale}. All distributions exhibit log-normal behavior. Furthermore, the distributions for the same data set and different years follow a universal curve with the same value of parameters $\mu$ and $\sigma$. The universal behavior is observed for the distribution of normalized log-rates as well, see Fig. \ref{fig:scale} (bottom panel). These results indicate that the growth of the social groups did not change due to the increased growth of members in systems. Furthermore, it implies that the growth is independent of the size of the whole data set.   

\section{Model \label{sec:model}}
The growth of social groups can not be explained by the simple rules of Gibrat's law. Previous research on group growth and longevity has shown that social connections with members of a group influence individual's choice to join that group \cite{kairam2012life, zheleva2009co}. Individuals' interests and the need to discover new content or activity also influence the diffusion of individuals between groups. Furthermore, social systems constantly grow since new members join every minute. The properties of the growth signal that describes the arrival of new members influence both dynamics of the system \cite{mitrovic2011quantitative, dankulov2015dynamics} and the structure of social interactions \cite{vranic2021growth}. The number of social groups in the social systems is not constant. They are constantly created and destroyed.

In Ref. \cite{zheleva2009co} authors propose the co-evolution model of the growth of social networks. In this model, the authors assume that the social system evolves through the co-evolution of two networks: a network of social contacts between members and a network of members' affiliations with groups. This model addresses the problem of the growth of social networks that includes both linking between members and social group formation. In this model, a member of a social system selects to join a group either through random selection or according to her social contacts. In the case of random selection, there is a selection preference for larger groups. If a member chooses to select a group according to her social contacts, the group is selected randomly from the list of groups with which her friends are already affiliated.

In Ref.\cite{zheleva2009co} authors demonstrate that mechanisms postulated in the model could reproduce the power-law distribution of group sizes observed for some social networks. However, as illustrated in Sec  \ Ref {sec:emp}, the distribution of group sizes in real systems is not necessarily power-law. Our rigorous empirical analysis shows that the distribution of social group sizes exhibits log-normal behavior. To fill the gap in understanding how social groups in the social system grow, we propose a model of group growth that combines random and social diffusion between groups but follows different rules than the co-evolution model \cite{zheleva2009co}.\\

\begin{figure}[!ht]
    \centering
   \includegraphics[scale=0.5]{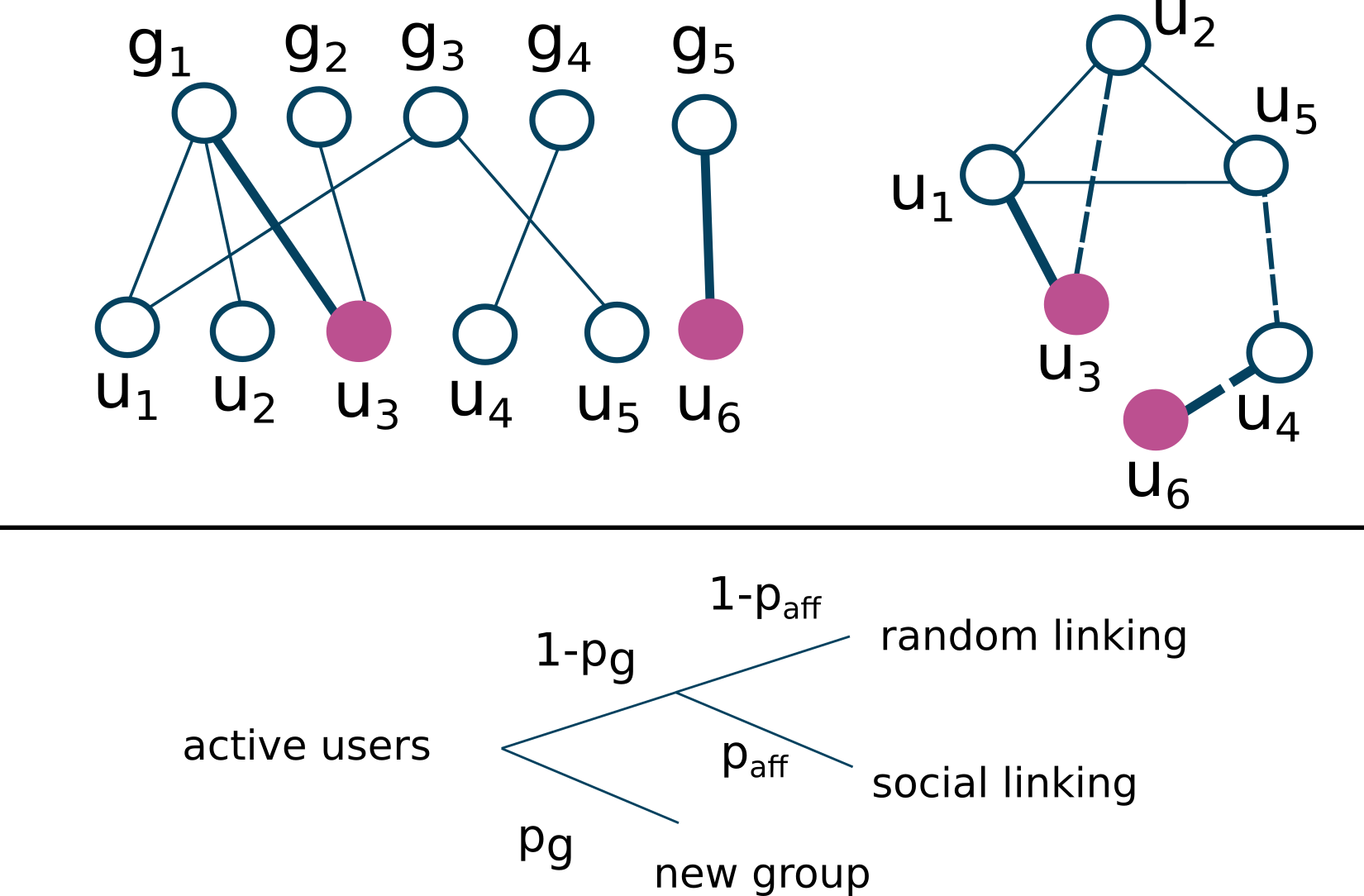}
  \caption{The top panel shows bipartite (member-group) and social (member-member) network. Filled nodes are active members, while thick lines are new links in this time step. In the social network dashed lines show that members are friends but still do not share same groups. The lower panel shows model schema. \textbf{Example:} member $u_6$ is a new member. First it will make random link  with node $u_4$, and then with probability $p_g$ makes new group $g_5$. With probability $p_a$ member $u_3$ is active, while others stay inactive for this time step. Member $u_3$ will with probability $1-p_g$ choose to join one of old groups and with probability $p_{aff}$ linking is chosen to be social. As its friend $u_2$ is member of group $g_1$, member $u_3$ will also join group $g_1$. Joining group $g_1$, member $u_3$ will make more social connections, in this case it is member $u_1$.}
    \label{fig:schema}
\end{figure}

Figure \ref{fig:schema} shows a schematic representation of our model. Similar to the co-evolution model \cite{zheleva2009co}, we represent a social system with two evolving networks, see Fig. \ref{fig:schema}. One network is a bipartite network that describes the affiliation of individuals to social groups $\mathcal{B}(V_{U}, V_{G}, E_{UG})$. This network consists of two partitions, members $V_{U}$ and groups $V_{G}$, and a set of links $E_{UG}$, where a link $e(u,g)$ between a member $u$ and a group $g$ represents the member's affiliation with that group. Bipartite network grows through three activities: the arrival of new members, the creation of new groups, and members joining groups. In bipartite networks, links only exist between nodes belonging to different partitions. However, as we explained above, social connections affect whether a member will join a certain group or not. In the simplest case, we could assume that all members belonging to a group are connected. However, previous research on this subject \cite{ smiljanic2017associative, backstrom2006group, zheleva2009co} has shown that the existing social connections of members in a social group are only a subset of all possible connections. For these reasons, we introduce another network $\mathcal{G}(V_{U},E_{UU})$ that describes social connections between members. The social network grows by adding new members to the set $V_{U}$ and creating new links between them. The member partition in bipartite network $\mathcal{B}(V_{U}, V_{G}, E_{UG})$ and set of nodes in members' network $\mathcal{G}(V_{U}, E_{UU})$ are identical.\\

For convenience, we represent the bipartite and social network of members with adjacency matrices $B$ and $A$. The element of the matrix $B_{ug}$ equals one if member $u$ is affiliated with group $g$, and zero otherwise. In matrix $A$, the element $A_{u_{1}u_{2}}$ equals one if members $u_{1}$ and $u_{2}$ are connected and zero otherwise. The neighborhood $\mathcal{N}_{u}$ of member $u$ is a set of groups with which the member is affiliated. On the other hand, the neighborhood $\mathcal{N}_{g}$ of a group $g$ is a set of members affiliated with that group. The size $S_{g}$ of set $\mathcal{N}_g$ equals to the size of the group $g$.\\

In our model, the time is discrete, and networks evolve through several simple rules. In each time step, we add $N_{U}(t)$ new members and increase the size of the set $V_{U}$. For each newly added member, we create the link to a randomly chosen old member in the social network $G$. This condition allows each member to perform social diffusion \cite{kairam2012life}, i.e., to select a group according to her social contacts. 
Not all members from setting $V_{U}$ are active in each time step. Only a subset of existing members is active in each time step. The activity of old members is a stochastic process determined by parameter $p_{a}$; every old member is activated with probability $p_{a}$. Old members are activated in this way, and new members make a set of active members $\mathcal{A}_{U}$ at time t.\\ 

The group partition $V_{G}$ grows through creating new groups. Each active member $u\in \mathcal{A}_{U}$ can decide with probability $p_{g}$ to create a new group or to join an already existing one with probability $1-p_{g}$. \\

If the active member $u$ decides that she will join an existing group, she first needs to choose a group. A member $u$ with probability $p_{aff}$ decides to select a group based on her social connections. For each active member, we look at how many social contacts she has in each group. The number of social contacts $s_{ug}$ that member $u$ has in the group $g$ equals the overlap of members affiliated with a group $g$ and social contacts of member $u$, and is calculated according to
\begin{equation}
    s_{ug}=\sum_{u_{1}\in \mathcal{N}_{g}}
    A_{uu_{1}} \label{eq1} \ .
\end{equation}
Member $u$ selects an old group $g$ to join according to probability $P_{ug}$ that is proportional to $s_{ug}$. Member-only considers groups with which it has no affiliation. However, if an active member decides to neglect her social contacts in the choice of the social group, she will select a random group from the set $V_{G}$ with which she is not yet affiliated. \\

After selecting the group $g$, a member joins that group, and we create a link in the bipartite network between a member $u$ and a group $g$. At the same time, the member selects $X$ members of a group $g$ which do not belong to her social circle and creates social connections with them. As a consequence of this action, we make $X$ new links in-network $\mathcal{G}$ between member $u$ and $X$ members from a group $g$.\\

The evolution of bipartite and social networks, and consequently growth of social groups, is determined by parameters $p_{a}$, $p_{g}$ and $p_{aff}$. Parameter $p_{a}$ determines the activity level of members and takes values between $0$ and $1$. Higher values of $p_{a}$ result in a higher number of active members and thus faster growth of the number of links in both networks and the size and number of groups. Parameter $p_{g}$ in combination with parameter $p_{a}$ determines the growth of the set $V_{G}$. $p_{g}=1$ means that members only create new groups, and the existing network consists of star-like subgraphs with members being central nodes and groups as leaves. On the other hand, $p_{g}=0$ means that there is no creation of new groups, and the bipartite network only grows through adding new members and creating new links between members and groups. \\

Parameter $p_{aff}$ determines the importance of social diffusion. $p_{aff}=0$ means that social connections are irrelevant, and the group choice is random. On the other hand, $p_{aff}=1$ means that only social contacts become important for group selection. \\

Several differences exist between the model presented in this work and the co-evolution model Ref. \cite{zheleva2009co}. In our model, $p_{aff}$ is constant and the same for all members. In the co-evolution model, this probability depends on members' degrees. The members are activated in our model with probability $p_{a}$. In contrast, in the co-evolution model, members are constantly active from the moment they are added to a set $V_{U}$ until they become inactive after time $t_{a}$. Time $t_{a}$ differs for every member and is drawn from an exponential distribution. In the co-evolution model, the number of social contacts members have within the group is irrelevant to its selection. On the other hand, in our model, members tend to choose groups more often in which there is a greater number of social contacts. While in our model, in the case of a random selection of a group, a member selects with equal probability a group that she is not affiliated with, in the co-evolution model, the choice of group is preferential.\\

\section{Results \label{sec:results}}

The distribution of group sizes produced by our and co-evolution models significantly differ. The distribution of group sizes in the co-evolution model is a power-law. Our model enables us to create groups with log-normal size distribution and expand classes of social systems that can be modeled. 

\subsection{Model properties}
First, we explore the properties of size distribution depending on parameters $p_{g}$ and $p_{aff}$, for the fixed value of activity parameter $p_{a}$ and constant number of members added in each step $N(t)=30$. We set the value of parameter $X$ to $25$ for all simulations presented in this work. Our detailed analysis of the results for different parameter values $X$ shows that these results are independent of their value.

\begin{figure}[!ht]
    \centering
    \includegraphics[width=0.8\linewidth]{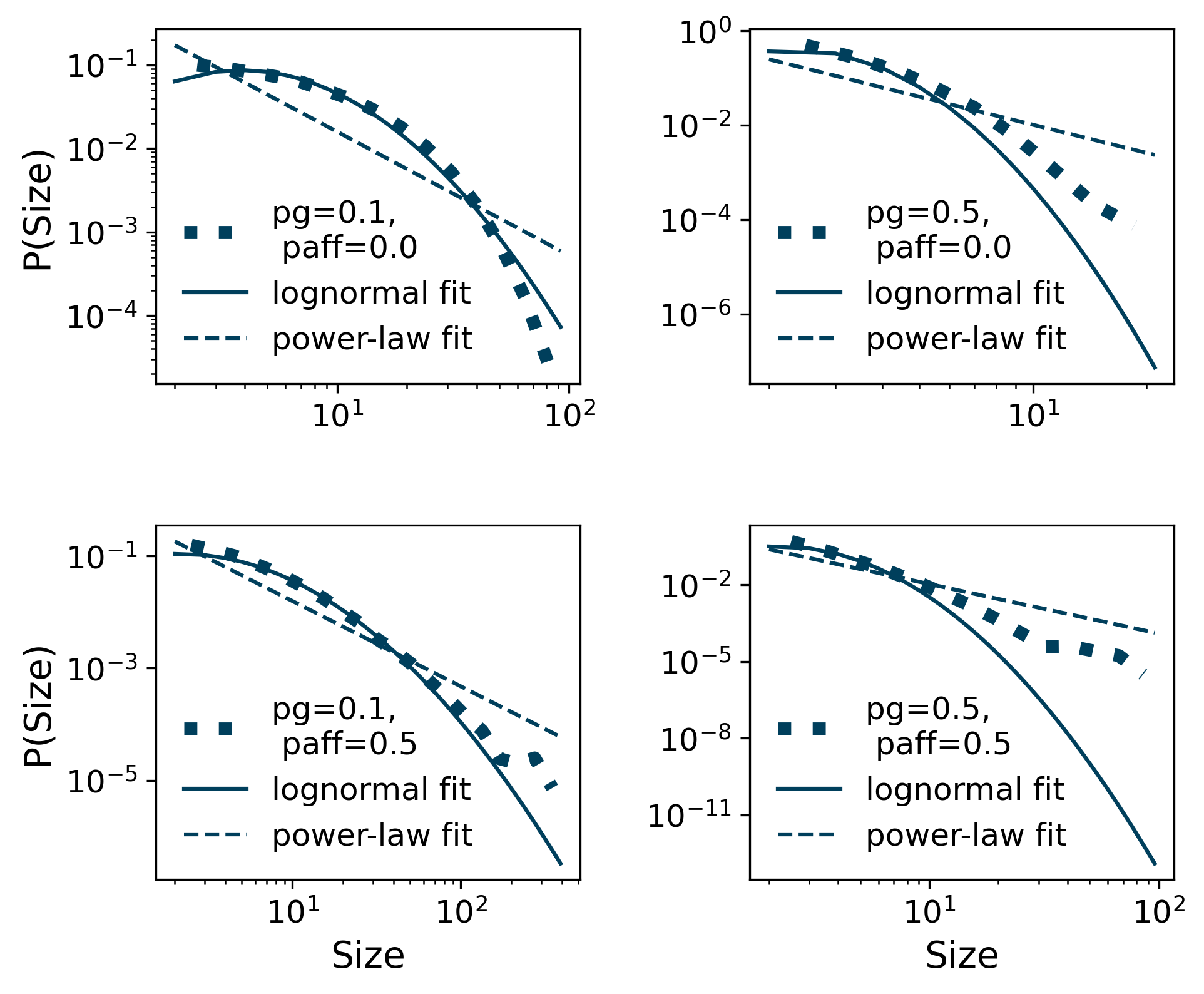}
    \caption{The distribution of sizes for different values of $p_{g}$ and $p_{aff}$ and constant $p_{a}$ and growth of the system. The combination of the values of parameters of $p_{g}$ and $p_{aff}$ determine the shape and the width of the distribution of group sizes. }
    \label{fig:n30}
\end{figure}

Figure \ref{fig:n30} shows some of the selected results and their comparison with power-law and log-normal fits. We see that values of both $p_g$ and $p_{aff}$ parameters, influence the type and properties of size distribution. For low values of parameter $p_{g}$, left column in Fig. \ref{fig:n30}, the obtained distribution is log-normal. The width of the distribution depends on $p_{aff}$. Higher values of $p_{aff}$ lead to a broader distribution.\\
As we increase $p_{g}$, right column Fig. \ref{fig:n30}, the size distribution begins to deviate from
log-normal distribution. The higher the value of parameter $p_{g}$, the total number of groups grows faster. For $p_{g}=0.5$, half of the active members in each time step create a group, and the number of groups increases fast. How members are distributed in these groups depends on the parameter $p_{aff}$ value. When $p_{aff}=0$, social connections are irrelevant to the group's choice, and members select groups randomly. The obtained distribution slightly deviates from log-normal, especially for large group sizes. In this case, large group sizes become more probable than in the case of the log-normal distribution. The non-zero value of parameter $p_{aff}$ means that the choice of a group becomes dependent on social connections. When a member chooses a group according to her social connections, larger groups have a higher probability of being affiliated with the social connections of active members, and thus this choice resembles preferential attachment. For these reasons, the obtained size distribution has more broad tail than log-normal distribution and begins to resemble power-law distribution.\\

The top panel of Fig. S3 in SI shows how the shape of distribution is changing with the value of parameter $p_{aff}$ and fixed values of $p_{a}=0.1$ and $p_{g}=0.1$. Preferential selection groups according to their size instead of one where a member selects a group with equal probability leads to a drastic change in the shape of the distribution, bottom panel Fig. S3 in SI. As is to be expected, the distribution of group sizes with preferential attachment follows power-law behavior.  

\subsection{Modeling real systems}
\begin{figure}[!ht]
    \centering
    \includegraphics[width=0.8\linewidth]{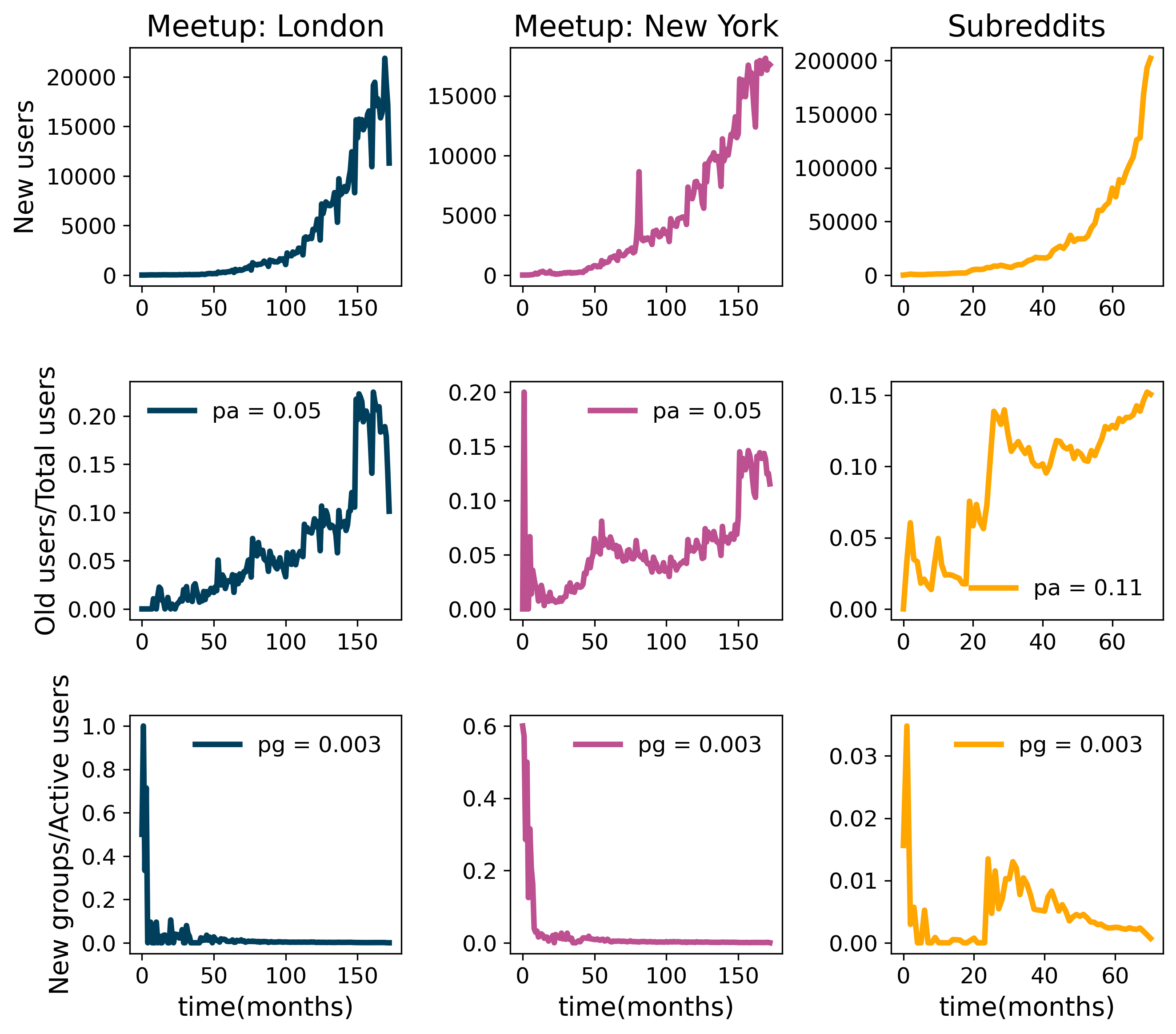}
    \caption{The time series of the number of new members (top panel). The time series of the ratio between several old active members and total members in the system (middle panel); its median value approximates the parameter $p_a$, the probability that the user is active. The bottom panel shows the time series of the ratio between new groups and active members; its median value approximates the probability that active users create a new group, parameter $p_g$.}
    \label{fig:fig5}
\end{figure}

The social systems do not grow at a constant rate. In Ref. \cite{vranic2021growth} authors have shown that features of growth signal influence the structure of social networks. For these reasons, we use the real growth signal from Meetup groups located in London and New York and Reddit to simulate the growth of the social groups in these systems. Figure \ref{fig:fig5} top panel shows the time series of the number of new members that join each of the considered systems each month. All three data sets have relatively low growth at the beginning, and then the growth accelerates as the system becomes more popular.\\ 

We also use empirical data to estimate $p_{a}$, $p_{g}$ and $p_{aff}$. The data can approximate the probability that old members are active $p_a$ and that new groups are created $p_g$. Activity parameter $p_{a}$ is the ratio between the number of old members active in month $t$ and the total number of members in the system at time $t$. Figure \ref{fig:fig5} middle row shows the variation of parameter $p_{a}$ during the considered time interval for each system. The value of this parameter fluctuates between $0$ and $0.2$ for London and New York based Meetup groups, while its value is between $0$ and $0.15$ for Reddit. To simplify our simulations, we assume that $p_{a}$ is constant in time and estimate its value as its median value during the $170$ months for Meetup and $80$ months for Reddit systems. For Meetup groups based in London and New York $p_{a}=0.05$, while Reddit members are more active on average and $p_{a}=0.11$ for this system.\\
Figure \ref{fig:fig5} bottom row shows the evolution of parameter $p_{g}$ for the considered systems. The $p_{g}$ in month $t$ is estimated as the ratio between the groups created in month $t$ $Ng_{new}(t)$ and the total number of groups in that month $Ng_{new}(t)+Ng_{old}(t)$, i.e., $p_{g}(t)=\frac{Ng_{new}(t)}{N_{new}(t)+N_{old}(t)}$. We see from Fig. \ref{fig:fig5} that $p_{g}(t)$ has relatively high values at the beginning of the system's existence. This is not surprising. Initially, these systems have a relatively small number of groups and often cannot meet the needs of the content of all their members. As the time passes, the number of groups and content scope within the system grows, and members no longer have a high need to create new groups. Figure \ Ref {fig:fig5} shows that $p_{g}$ fluctuates less after the first few months, and thus we again assume that $p_{g}$ is constant in time and set its value to the median value during 170 months for Meetup and 80 months for Reddit. For all three systems $p_{g}$ has the value of $0.003$.\\ 
The affiliation parameter $p_{aff}$ cannot estimate directly from the empirical data. For these reasons, we simulate the growth of social groups for each data set with the time series of new members obtained from the real data and estimated values of parameters $p_a$ and $p_g$, while we vary the value of $p_{aff}$. We compare the distribution of group sizes obtained from simulations for different values of $p_{aff}$ with ones obtained from empirical analysis using Jensen Shannon (JS) divergence. The JS divergence \cite{jsdivergence} between two distributions $P$ and $Q$ is defined as 
\begin{equation}
    JS(P, Q) = H\left(\frac{P+Q}{2}\right) - \frac{1}{2}\left(H(P)+H(Q)\right) \label{eq2}
\end{equation}
where $H(p)$ is Shannon entropy $H(p)=\sum_x p(x)log(p(x)$. The JS divergence is symmetric and if $P$ is identical to $Q$, $JS=0$. The smaller the value of JS divergence, the better is the match between empirical and simulated group size distributions. Table \ref{tab:table} shows the value of JS divergence for all three data sets. We see that for London based Meetup groups the affiliation parameter is $p_{aff}=0.5$, for New York groups $p_{aff}=0.4$, while the affiliation parameter for Reddit $p_{aff}=0.8$. Our results show that social diffusion is important in all three data sets. However, Meetup members are more likely to join groups at random, while for the Reddit members their social connections are more important when it comes to choice of the subreddit.

\begin{table}[h]
\centering
\begin{tabular}{|c|c|c|c|}
\hline
$p_{aff}$ & JS cityLondon   & JS cityNY       & JS reddit2012    \\ \hline
0.1  & 0.0161          & 0.0097          & 0.00241          \\ \hline
0.2  & 0.0101          & 0.0053          & 0.00205          \\ \hline
0.3  & 0.0055          & 0.0026          & 0.00159          \\ \hline
0.4  & 0.0027          & \textbf{0.0013} & 0.00104          \\ \hline
0.5  & \textbf{0.0016} & 0.0015          & 0.00074          \\ \hline
0.6  & 0.0031          & 0.0035          & 0.00048          \\ \hline
0.7  & 0.0085          & 0.0081          & 0.00039          \\ \hline
0.8  & 0.0214          & 0.0167          & \textbf{0.00034} \\ \hline
0.9  & 0.0499          & 0.0331          & 0.00047          \\ \hline
\end{tabular}
\caption{Jensen Shannon divergence between group sizes distributions from model and data. In model we vary affiliation parameter $p_{aff}$ and find its optimal value (bold text).}
\label{tab:table}
\end{table}

\begin{figure}[!ht]
    \centering
    \includegraphics[width=0.8\linewidth]{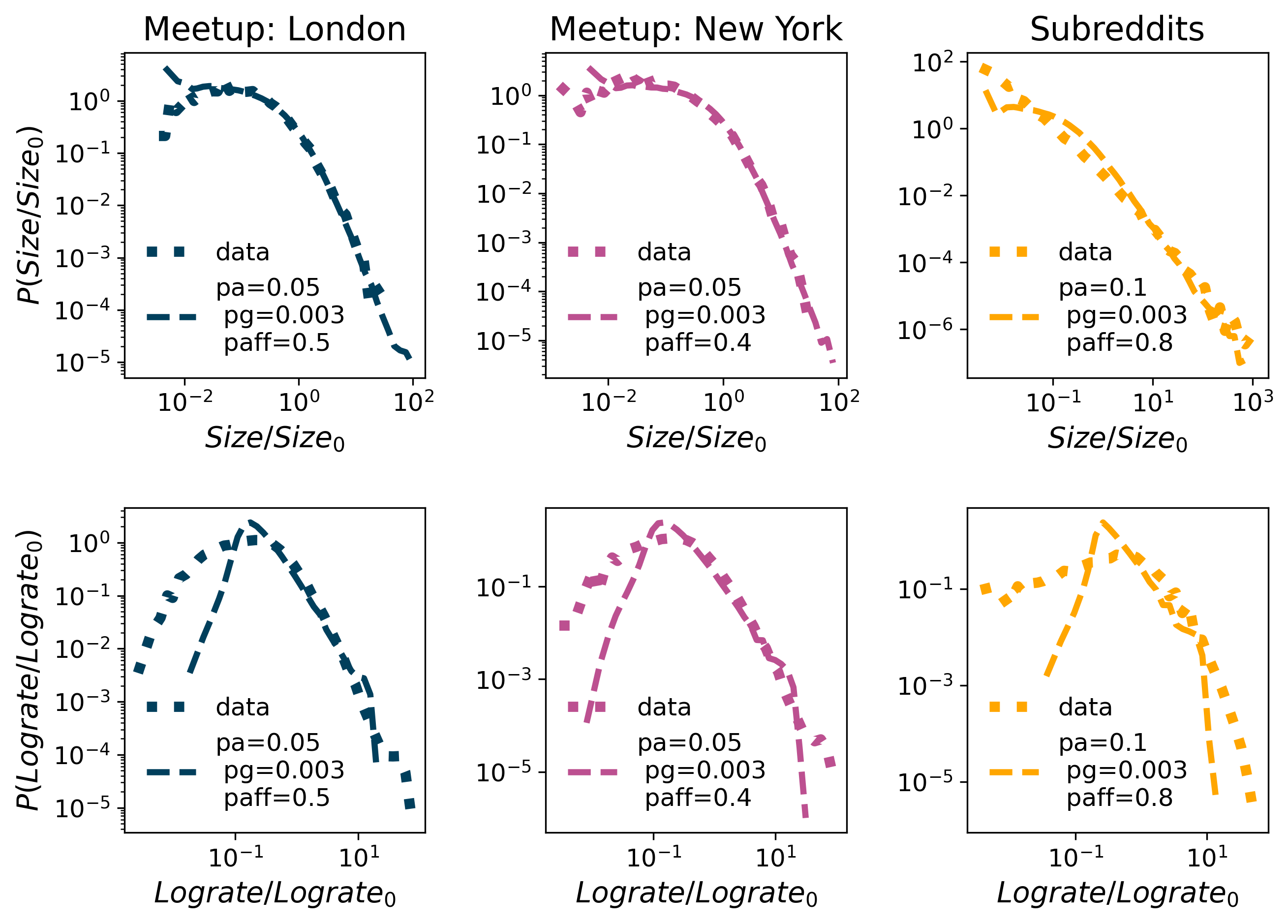}
    \caption{The comparison between empirical and simulation distribution for group sizes (top panel) and log-rates (bottom panel).}
    \label{fig:fig6}
\end{figure}

Figure \ Ref {fig:fig6} compares the empirical and simulation distribution of group sizes for considered systems. We see that empirical distributions for Meetup groups based in London and New York are well reproduced by the model and chosen values of parameters. In the case of Reddit, the distribution is broad, and the model reproduces the tail of the distribution well. Figure S2 and Table S2 in SI confirm that the distribution of group sizes follow a log-normal distribution.\\
The bottom row of Fig. \ref{fig:fig6} shows the distribution of logarithmic values of growth rates of groups obtained from empirical and simulated data. We see that the tails of empirical distributions for all three data sets are well emulated by the ones obtained from the model. The deviations we observe are the most likely consequence of using median values of parameters $p_{a}$, $p_{g}$, and $p_{aff}$.\\

\section{Discussion and conclusions \label{sec:con}}

The results of empirical analysis show that there are universal growth rules that govern
the growth of social systems. We analysed the growth of social groups for three data sets,
Meetup groups located in London and New York and Reddit. We showed that the
distribution of group sizes has log-normal behaviour. The empirical distributions of
normalised sizes of groups created in different years in a single system fall on top of
each other, following the same log-normal distributions. Due to a limited data availability, we only study three data sets which may affect the generality of our results. However, the substantial differences between Reddit and Meetup social systems when it comes to their popularity, size and purpose, demonstrate that observed growth patterns are universal. 

Even though the log-normal distribution of group sizes can originate from the proportional growth model, Gibrat law, we show that it does not apply to the growth of online social groups. The monthly growth rates are log-normally distributed and dependent on the size of a group. Gibrat law was proposed to describe the growth of various socio-economical systems, including the cities and firms. Recent studies
showed that the growth of cities and firms \cite{mansfield1962entry,stanley1996scaling,barthelemy2019statistical} goes beyond Gibrat law. Still, our
findings confirm the existence of universal growth patterns, indicating the presence of
the general law in the social system's growth.

The model proposed in Ref. \cite{zheleva2009co} is able to produce only power-law distributions of group sizes. However, our empirical analysis shows that these distributions can also have a log-normal behavior. Thus, we propose a new model that emulate log-normal distributions. The analysed groups grow through two mechanisms \cite{zheleva2009co}: members
join a group that is chosen according to their interests or by social relations with the
group's members. The number of members in the system is growing as well as the number of
groups. While the processes that govern the growth of social groups are the same, their
importance varies among the systems. The distributions for Meetup groups located in
the London and New York have similar log-normal distribution parameter values, while
for Reddit, the distribution is broader. Numerical simulations further confirm these
findings. Different modalities of interactions between their members can explain the
observed differences.

Meetup members need to invest more time and resources to interact with their peers. The events are localised in time and space, and thus the influence of peers in selecting another social group may be limited. On the other hand, Reddit members do not have these limitations. The interactions are online, asynchronous, and thus not limited in time. The influence of peers in choosing new subreddits and topics thus becomes more important. The values of $p_{aff}$ parameters for Meetup and Reddit imply that social connections in diffusion between groups are more critical in Reddit than in Meetup.

The results presented in this paper contribute to our knowledge of the growth of
socio-economical systems. The previous study analysed the social systems in which
size distributions follow the power-law, which is the consequence of a preferential choice of groups during the random diffusion of members. Our findings show that preferential selection of groups during social diffusion and uniform selection during random diffusion result in log-normal distribution of groups sizes. Furthermore, we show that broadness of the distribution depends on the involvement of social diffusion in the growth process. Our model increases the number of
systems that can be modelled and help us better understand the growth and
segmentation of social systems and predict their evolution.

\section*{References}
\bibliography{references.bib}
\bibliographystyle{unsrt}

\end{document}


\title[]{Supplementary Information: Universal growth of social groups: empirical analysis and modeling}
\section{Distributions fit}

We compute the log-likelihood ratio $R$, and $p$-value between different distributions and log-normal fit \cite{clauset2009power} to determine the best fit for the group size distributions. Distribution with a higher likelihood is a better fit. The log-likelihood ratio R then has a positive or negative value, indicating which distribution represents a better fit. To choose between two distributions, we need to calculate p-value,  to be sure that R is sufficiently positive or negative and that it is not the result of chance fluctuation from the result that is close to zero. If the p-value is small, $p<0.1$, it is unlikely that the sign of R is the chance of fluctuations, and it is an accurate indicator of which model fits better. \\

\begin{figure}[!h]
    \centering
    \includegraphics[width=\linewidth]{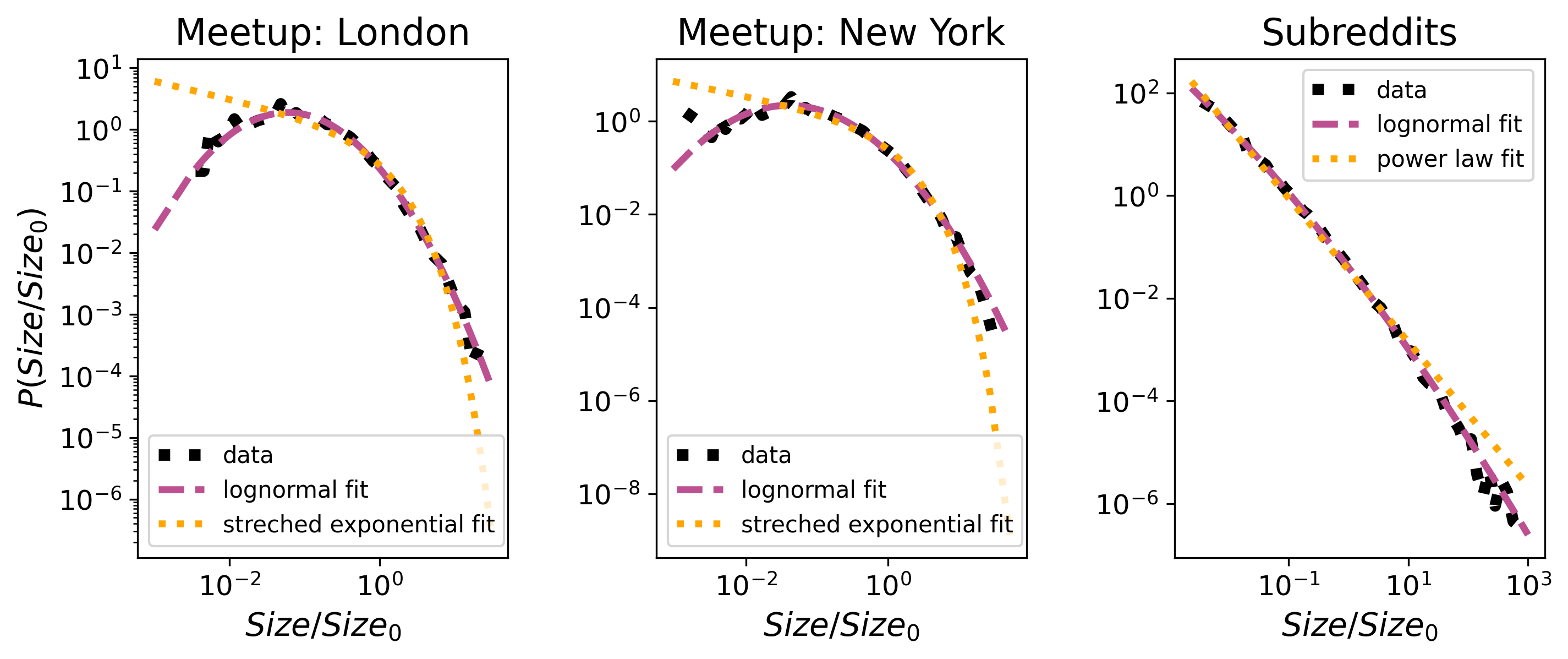}
    \caption{The comparison between log-normal and stretched exponential fit to London and NY data,  and between log-normal and power law for Subreddits. The parameters for log-normal fits are 1) for city London $\mu=-0.93$ and $\sigma = 1.38$, 2) for city NY $\mu=-0.99$ and $\sigma = 1.49$, 3) for Subreddits $\mu=-5.41$ and $\sigma = 3.07$.  }
    \label{fig:fitdata}
\end{figure}

Table \ref{tab:fit-data} summarizes the findings for empirical data on group size distributions from Meetup groups in London, Meetup groups in New York and Reddit. Using the maximum likelihood method, we obtain the parameters of the distributions \cite{powerlaw}. The results indicate that log-normal distribution is the best fit for all three systems.  Figure \ref{fig:fitdata} shows the distributions of empirical data as well as log-normal fit on data. For Meetup data, we present fit on stretched exponential distribution, which very well fits a large portion of data. For subreddits, distribution is broad and, potentially, resembles power-law. Still, log-normal distribution is a more suitable fit.

\begin{table}[!h]
\centering
\caption{The likelihood ratio R and p-value between different candidates and \textbf{lognormal} distribution for fitting the distribution of \textbf{groups sizes} of Meetup groups in London, New York and in Reddit. According to these statistics, the lognormal distribution represents the best fit for all communities. \\ }
\begin{tabular}{|c||cc||cc||cc|}
\hline
\multirow{2}{*}{\begin{tabular}[c]{@{}c@{}}distribution \end{tabular}} & \multicolumn{2}{c||}{\begin{tabular}[c]{@{}c@{}}Meetup\\ city London\end{tabular}} & \multicolumn{2}{c||}{\begin{tabular}[c]{@{}c@{}}Meetup\\ city NY\end{tabular}} & \multicolumn{2}{c|}{Reddit}                    \\ \cline{2-7} 
                                                                                       & \multicolumn{1}{c|}{R}                             & p                            & \multicolumn{1}{c|}{R}                           & p                          & \multicolumn{1}{c|}{R}         & p             \\ \hline \hline \hline
exponential                                                                            & \multicolumn{1}{c|}{-8.64e2
}                       & 8.11e-32                     & \multicolumn{1}{c|}{-8.22e2}                     & 6.63e-26                   & \multicolumn{1}{c|}{-3.85e4} & 1.54e-100     \\ \hline
\begin{tabular}[c]{@{}c@{}}stretched \\ exponential\end{tabular}                       & \multicolumn{1}{c|}{-3.01e2}                       & 1.00e-30                      & \multicolumn{1}{c|}{-1.47e2}                     & 7.78e-8                    & \multicolumn{1}{c|}{-7.97e1}    & 5.94e-30      \\ \hline
power law                                                                              & \multicolumn{1}{c|}{-4.88e3}                      & 0.00                         & \multicolumn{1}{c|}{-4.57e3}                    & 0.00                       & \multicolumn{1}{c|}{-9.39e2}   & 4.48e-149 \\ \hline
\begin{tabular}[c]{@{}c@{}}truncated \\ power law\end{tabular}                         & \multicolumn{1}{c|}{-2.39e3}                      & 0.00                         & \multicolumn{1}{c|}{-2.09e3}                    & 0.00                       & \multicolumn{1}{c|}{-5.51e2}   & 2.42e-56      \\ \hline
\end{tabular}
\label{tab:fit-data}
\end{table}

\begin{figure}[!h]
    \centering
    \includegraphics[width=\linewidth]{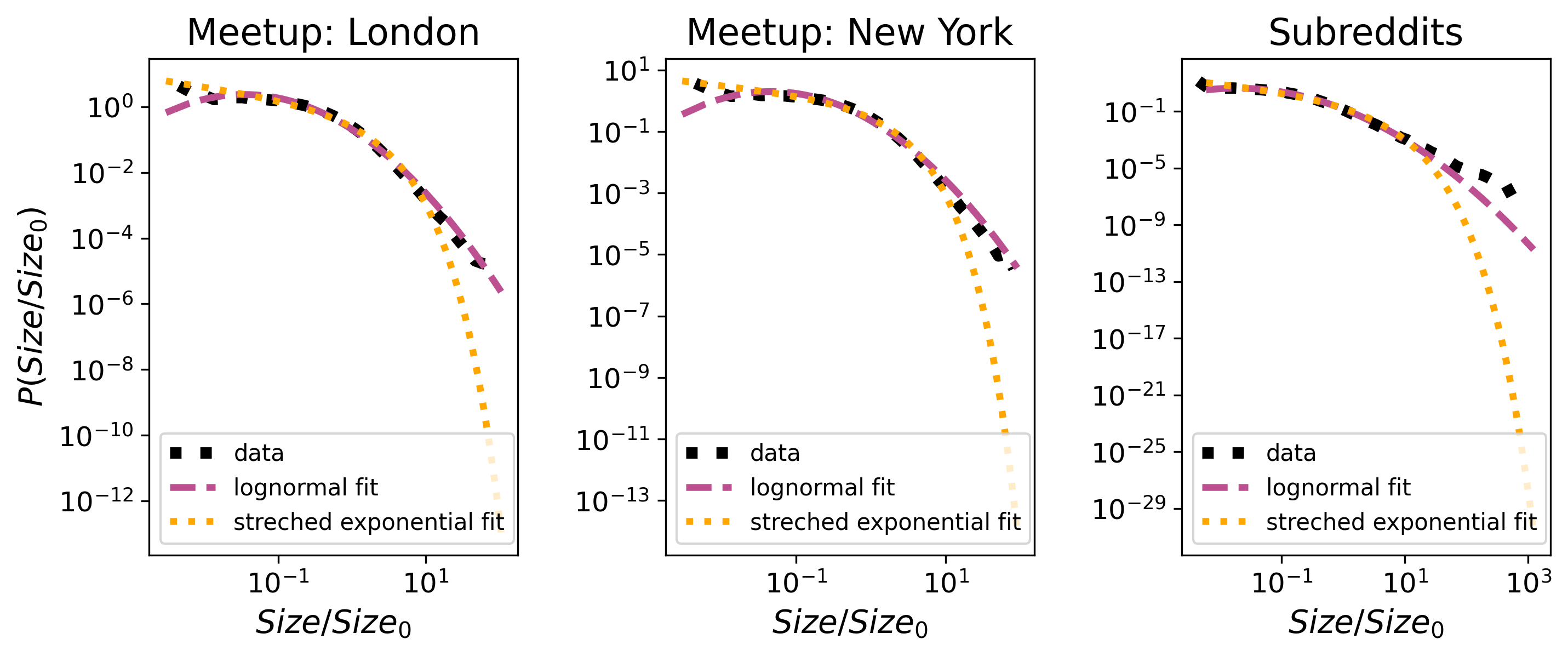}
    \caption{The comparison between lognormal and stretched exponential fit to simulated group sizes distributions. The parameters for log-normal fits are 1) for city London $\mu=-0.97$ and $\sigma = 1.43$ , 2) for city NY $\mu=-0.84$ and $\sigma = 1.38$ , 3) for Subreddits $\mu=-1.63$ and $\sigma = 1.53$. }
    \label{fig:fit_model}
\end{figure}

We use the same methods to estimate the fit for simulated group size distributions on Meetup groups in London, New York, and Subreddits. Table \ref{tab:fit_model} shows the results of the log-likelihood ratio R and $p$-value between different distributions. We conclude that log-normal distribution is most suitable for simulated group size distributions. Plotting log-normal and stretched exponential fit on data, Fig. \ref{fig:fit_model} we confirm our observations.  
\begin{table}[!h]
\centering
\caption{The likelihood ratio R and p-value between different candidates and \textbf{lognormal}
distribution for fitting the distribution of \textbf{simulated group sizes} of Meetup groups in London, New York and Reddit. According to these statistics, the lognormal distribution
represents the best fit for all communities. \\}
\begin{tabular}{|c||cc||cc||cc|}
\hline
\multirow{2}{*}{\begin{tabular}[c]{@{}c@{}}distribution \end{tabular}} & \multicolumn{2}{c||}{\begin{tabular}[c]{@{}c@{}}Meetup\\ city London\end{tabular}} & \multicolumn{2}{c||}{\begin{tabular}[c]{@{}c@{}}Meetup\\ city NY\end{tabular}} & \multicolumn{2}{c|}{Reddit}                \\ \cline{2-7} 
                                                                                       & \multicolumn{1}{c|}{R}                              & p                           & \multicolumn{1}{c|}{R}                            & p                         & \multicolumn{1}{c|}{R}         & p         \\ \hline \hline \hline
exponential                                                                            & \multicolumn{1}{c|}{-6.27e4}                      & 0.00                        & \multicolumn{1}{c|}{-5.11e4}                    & 0.00                      & \multicolumn{1}{c|}{-1.26e5} & 7.31e-125 \\ \hline
\begin{tabular}[c]{@{}c@{}}stretched\\ exponential\end{tabular}                        & \multicolumn{1}{c|}{-1.01e4}                      & 1.96e-287                    & \multicolumn{1}{c|}{-6.69e3}                     & 1.46e-93                  & \multicolumn{1}{c|}{-1.39e4} & 0.00      \\ \hline
power law                                                                              & \multicolumn{1}{c|}{-2.29e5}                     & 0.00                        & \multicolumn{1}{c|}{-3.73e5}                   & 0.00                      & \multicolumn{1}{c|}{-4.38e4} & 0.00      \\ \hline
\begin{tabular}[c]{@{}c@{}}truncated\\ power law\end{tabular}                          & \multicolumn{1}{c|}{-9.28e4}                      & 0.00                        & \multicolumn{1}{c|}{-1.55e5}                   & 0.00                      & \multicolumn{1}{c|}{-9.12e4} & 0.00      \\ \hline
\end{tabular}
\label{tab:fit_model}
\end{table}
\clearpage

\section{The model for social groups growth}
\begin{figure}[!h]
    \centering
    \includegraphics[width=\linewidth]{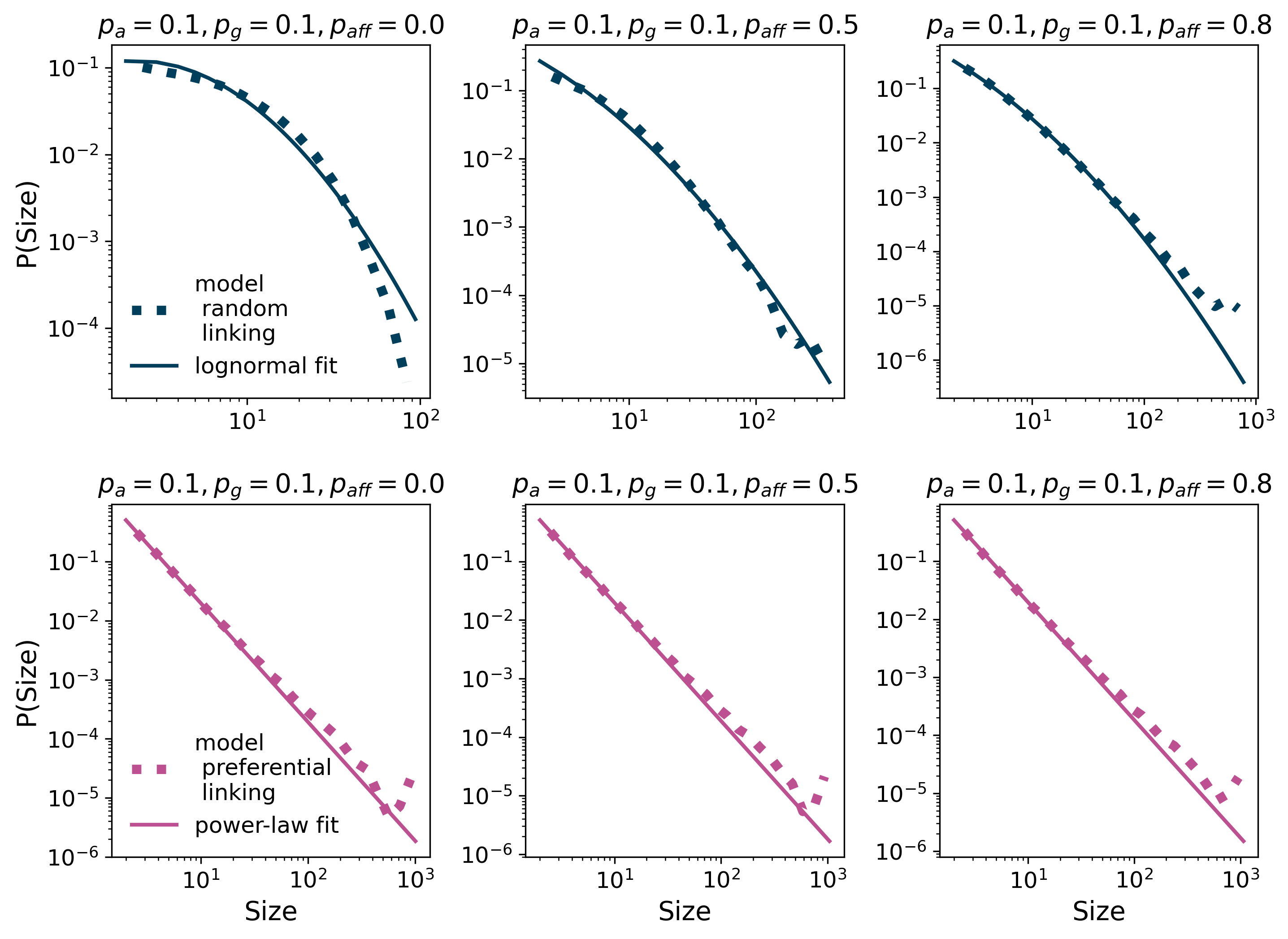}
    \caption{Groups sizes distributions for groups model, where at each time step the constant number of users arrive, $N=30$ and old users are active with probability $p_a=0.1$. Active users make new groups with probability $p_g=0.1$, while we vary affiliation parameter $p_{aff}$. With probability, $1-p_{aff}$, users choose a group randomly. The group sizes distribution (top row) is described with a log-normal distribution. With higher affiliation parameter, $p_{aff}$, distribution has larger width. The bottom row presents the case where with probability $1-p_{aff}$ users have a preference toward larger groups. For all values of parameter $p_{aff}$, we find the power-law group sizes distribution.}
     \label{fig:model_comp}
\end{figure}

In the groups growth model, at each step, new users join the network, while old users are active with probability $p_a$. Active users can create new group with probability $p_g$. Otherwise, with probability $p_{aff}$, they perform diffusion linking. With probability, $1-p_{aff}$ users join a random group. Figure \ref{fig:model_comp}, top row, shows that group sizes distributions follow log-normal distribution. The affiliation parameter $p_{aff}$ influences the width of distributions, so for larger $p_{aff}$, we find larger groups in the network.   If, instead of random linking, users with probability  $1-p_{aff}$, choose to join to larger groups, group sizes distribution change significantly. Similar to affiliation model \cite{zheleva2009co}, group sizes have power-law distribution, see bottom row on Figure \ref{fig:model_comp}.
\clearpage

\section*{References}
\bibliography{references.bib}
\bibliographystyle{unsrt}